\documentstyle[preprint,eqsecnum,aps]{revtex}
\begin{document}

\def\l{{\lambda}}

\def\eg{{\it e.\ g.}}\def\ie{{\it i.\ e.}}\def\etc{{\it etc.}}
\def\vs{{\it vs.}}\def\vv{{\it vice versa}}\def\ea{{\it et al.}}
\def\apr{{\it a priori}}\def\apo{{\it a posteriori}}

\newcount\refnum\refnum=0  
\def\refi{\smallskip\global\advance\refnum by 1\item{\the\refnum.}}

\newcount\fignum\fignum=0  
\def\figi{\global\advance\fignum by 1 Fig.~\the\fignum}

\newcount\eqnum \eqnum=0  
\def\eqnoi{\global\advance\eqnum by 1\eqno(\the\eqnum)}
\def\eqnai{\global\advance\eqnum by 1\eqno(\the\eqnum {\rm a})}
\def\back#1{{\advance\eqnum by-#1 Eq.~(\the\eqnum)}}
\def\last{Eq.~(\the\eqnum)~}                   

\def\pd#1#2{{\partial #1\over\partial #2}}      
\def\p2d#1#2{{\partial^2 #1\over\partial #2^2}} 
\def\pnd#1#2#3{{\partial^{#3} #1\over\partial #2^{#3}}} 
\def\td#1#2{{d #1\over d #2}}      
\def\t2d#1#2{{d^2 #1\over d #2^2}} 
\def\tnd#1#2#3{{d^{#3} #1\over d #2^{#3}}} 
\def\av#1{\langle #1\rangle}                    
\def\anv#1#2{\langle #1^{#2}\rangle}            


\def\acp #1 #2 #3 {{\sl Adv.\ Chem.\ Phys.} {\bf #1}, #2 (#3)}
\def\cp #1 #2 #3 {{\sl Chem.\ Phys.} {\bf #1}, #2 (#3)}
\def\eul #1 #2 #3 {{\sl Europhys.\ Lett.} {\bf #1}, #2 (#3)}
\def\jcp #1 #2 #3 {{\sl J.\ Chem.\ Phys.} {\bf #1}, #2 (#3)}
\def\jdep #1 #2 #3 {{\sl J.\ de Physique I} {\bf #1}, #2 (#3)}
\def\jdepl #1 #2 #3 {{\sl J. de Physique Lett.} {\bf #1}, #2 (#3)}
\def\jetp #1 #2 #3 {{\sl Sov.\ Phys.\ JETP} {\bf #1}, #2 (#3)}
\def\jetpl #1 #2 #3 {{\sl Sov. Phys.\ JETP Letters} {\bf #1}, #2 (#3)}
\def\jmp #1 #2 #3 {{\sl J. Math. Phys.} {\bf #1}, #2 (#3)}
\def\jpa #1 #2 #3 {{\sl J. Phys.\ A} {\bf #1}, #2 (#3)}
\def\jpchem #1 #2 #3 {{\sl J. Phys.\ Chem.} {\bf #1}, #2 (#3)}
\def\jpsj #1 #2 #3 {{\sl J. Phys.\ Soc. Jpn.} {\bf #1}, #2 (#3)}
\def\jsp #1 #2 #3 {{\sl J. Stat.\ Phys.} {\bf #1}, #2 (#3)}
\def\nat #1 #2 #3 {{\sl Nature} {\bf #1}, #2 (#3)}
\def\pA #1 #2 #3 {{\sl Physica A} {\bf #1}, #2 (#3)}
\def\pD #1 #2 #3 {{\sl Physica D} {\bf #1}, #2 (#3)}
\def\pla #1 #2 #3 {{\sl Phys.\ Lett. A} {\bf #1}, #2 (#3)}
\def\plb #1 #2 #3 {{\sl Phys.\ Lett. B} {\bf #1}, #2 (#3)}
\def\pr #1 #2 #3 {{\sl Phys.\ Rev.} {\bf #1}, #2 (#3)}
\def\pra #1 #2 #3 {{\sl Phys.\ Rev.\ A} {\bf #1}, #2 (#3)}
\def\prb #1 #2 #3 {{\sl Phys.\ Rev.\ B} {\bf #1}, #2 (#3)}
\def\pre #1 #2 #3 {{\sl Phys.\ Rev.\ E} {\bf #1}, #2 (#3)}
\def\prl #1 #2 #3 {{\sl Phys.\ Rev.\ Lett.} {\bf #1}, #2 (#3)}
\def\prslA #1 #2 #3 {{\sl Proc.\ R.\ Soc.\ London, Ser.\ A} {\bf #1}, #2 (#3)}
\def\rmp #1 #2 #3 {{\sl Rev.\ Mod.\ Phys.} {\bf #1}, #2 (#3)}
\def\rpp #1 #2 #3 {{\sl Rep.\ Progr.\ Phys.} {\bf #1}, #2 (#3)}
\def\sci #1 #2 #3 {{\sl Science} {\bf #1}, #2 (#3)}
\def\sciam #1 #2 #3 {{\sl Scientific American} {\bf #1}, #2 (#3)}
\def\usp #1 #2 #3 {{\sl Sov.\ Phys.\ Usp.} {\bf #1}, #2 (#3)}
\def\zpb #1 #2 #3 {{\sl Z. Phys.\ B} {\bf #1}, #2 (#3)}
\def\zpc #1 #2 #3 {{\sl Z. Phys.\ Chem.} {\bf #1}, #2 (#3)}


\centerline{\bf Kinetics of Aggregation-Annihilation Processes}
\bigskip\bigskip
\centerline{\bf E.~Ben-Naim$^1$ and P.~L.~Krapivsky$^2$}
\medskip
\centerline{$^1$The James Franck Institute}
\centerline{The University of Chicago, Chicago, IL 60637}
\medskip
\centerline{$^2$Center for Polymer Studies and Department of Physics}
\centerline{Boston University, Boston, MA 02215}

\begin{abstract}

We investigate the kinetics of many-species systems with aggregation
of similar species clusters and annihilation of opposite species
clusters.  We find that the interplay between aggregation and
annihilation leads to rich kinetic behaviors and unusual conservation
laws.  On the mean-field level, an exact solution for the cluster-mass
distribution is obtained.  Asymptotically, this solution exhibits a
novel scaling form if the initial species densities are the same while
in the general case of unequal densities the process approaches single
species aggregation.  The theoretical predictions are compared with
numerical simulations in 1D, 2D, and 3D.  Nontrivial growth exponents
characterize the mass distribution in one dimension.

\end{abstract}

\pacs{PACS numbers: 02.50.-r, 05.40.+j, 82.20.-w}

\vfil\eject

Irreversible aggregation and annihilation processes occur in many
natural phenomena \cite{smoke,meak}.  The kinetics of each of these
processes is well understood both on the mean-field level which
provides an accurate description when fluctuations in reactant
densities can be ignored \cite{smol,smoke,ernst} and in the
fluctuation-dominated regime which occurs in low-dimensional systems,
see \cite{zel}, and references therein.  It was found in particular
that the cluster-mass distribution in aggregating systems typically
approaches a universal scaling form \cite{meak,ernst}.

In this work we investigate the competition between aggregation and
annihilation, which gives rise to a surprisingly rich kinetic
behavior. Our model is the simplest implementation of the two
underlying processes, aggregation and annihilation: Similar species
clusters combine to form larger clusters while dissimilar clusters
combine to form an inert substitute.  It is worth noting that when
the number of species $n$ diverges, $n\to\infty$, our model is
equivalent to single-species
annihilation, while in the other extreme, $n=1$, single-species
aggregation is recovered. Hence, the process is well suited for
investigating the interplay between aggregation and annihilation.  We find
that the exponents characterizing the kinetics are nonuniversal in
that they may depend on the reaction rates as well as the initial
conditions. Still, there is a basic scaling form which describes the
process in the long-time limit.

We shall study our model both in the mean-field limit and in the
diffusion-controlled limit for $d=1,2$ and 3. The mean-field approach
to the binary reaction process assumes that the reaction proceeds with
a rate proportional to the product of the reactants densities. Thus
the mean-field approximation neglects spatial correlations and
therefore typically holds in dimensions larger than some critical
dimension $d_c$.  For $n$-species pure annihilation processes, it was
suggested \cite{bAR} that $d_c=4(n-1)/(2n-3)$, the result which turns
into rigorously established values \cite{leb} of the critical
dimension for two- and single-species annihilation, $d_c=4$ and 2,
respectively (the latter case corresponds to $n=\infty$).  For pure
aggregation processes, the critical dimension depends on the details
of the reaction events and on the relation between the diffusion
coefficient and the mass of the cluster \cite{vD}.  Numerically, we
investigate the particle coalescence model (PCM) in which clusters
occupy single lattice sites and perform nearest neighbor hoping with
the diffusion coefficient independent of the mass.  For the PCM, it
is known that $d_c=2$ \cite{kang}. Since the aggregation-annihilation
model interpolates between single-species aggregation and
single-species annihilation processes, both of which belong to the
same universality class, it is natural to expect that the critical
dimension remains the same, $d_c=2$.  We confirm the mean-field
predictions above this critical dimension numerically.  In
one dimension, spatial correlations are relevant asymptotically. A
simple heuristic argument provides the exact asymptotic behavior of
the concentration and a good approximation for the growth of the
typical mass.

Following the above discussion, similar species aggregation is
described by the binary reaction scheme
$$
A_i+A_j\to A_{i+j}, \quad
B_i+B_j\to B_{i+j},
\eqnoi
$$
where $A_i$ denotes a cluster consisting of $i$ monomers of species
$A$, and similarly for $B_i$. The two-species case contains the
generic many-species behavior. Therefore, we shall focus on the two-species
situation and cite the many-species results if appropriate.
Annihilation between dissimilar clusters reads
$$
A_i+B_j\to {\rm inert}.
\eqnoi
$$
Thus, we assume that dissimilar clusters completely annihilate
independent of their masses. In some situations it may be more
reasonable to assume a partial annihilation, where the monomer
difference number $i-j$ is conserved \cite{paul1,paul2,blum}.

Denote by $a_k(t)$ and $b_k(t)$ the densities of $A$- and $B$-clusters,
consisting of $k$ monomers. Then the mean-field rate equations
for the two-species aggregation-annihilation process read
$$
\dot a_k=\sum_{i+j=k} a_ia_j-2a_k(a+b),\quad
\dot b_k=\sum_{i+j=k} b_ib_j-2b_k(a+b).
\eqnoi
$$
Here $a(t)$ and $b(t)$ denote the total densities of $A$- and
$B$-clusters, $a=\sum _{k\ge 1}a_k$ and $b=\sum _{k\ge 1}b_k$,
and the overdot denotes the time derivative.  In writing \last\ we
have assumed that the aggregation rate equals the annihilation
rate. We set this constant rate to unity without loss of generality.

Summing up \last\ we obtain
$$
\dot a=-a^2-2ab, \quad
\dot b=-b^2-2ab.
\eqnoi
$$
For the symmetric initial conditions, $a(0)=b(0)$,
we get $\dot a=-3a^2$ which is immediately solved to find
$$
a(t)=b(t)={a(0)\over 1+3a(0)t}.
\eqnoi
$$
For the asymmetric initial conditions, $a(0)>b(0)$,
it is helpful to rewrite \back1\ in terms of $u\equiv (a+b)/(a-b)$
and $v\equiv a-b$. This gives
$$
\dot u=-{1\over 2}v(u^2-1), \quad
\dot v=-uv^2.
\eqnoi
$$
Expressing $v$ as a function of $u$ yields $dv/du=2uv/(u^2-1)$,
and as a result
$$
v=K(u^2-1), \quad
K={[a(0)-b(0)]^3\over 4a(0)b(0)}.
\eqnoi
$$
Inserting \last\ into  \back1, we obtain a closed
equation for $u\equiv u(t)$ with the solution
$$
{2u\over u^2-1}-{2u(0)\over u^2(0)-1}
+\ln\left[{u-1\over u+1}~{u(0)+1\over u(0)-1}\right]=2Kt.
\eqnoi
$$
The densities $a(t)$ and $b(t)$ of the total number of $A$- and
$B$-clusters, $a=v(u+1)/2$ and $b=v(u-1)/2$, cannot be expressed
as explicit functions of $t$. However, asymptotically they exhibit
simple power-law behaviors
$$
a\simeq t^{-1}, \quad
b\simeq a(0)b(0)[a(0)-b(0)]^{-3}t^{-2}
\quad {\rm for}\quad t\to\infty.
\eqnoi
$$
This indicates that the majority and the minority species evolve very
differently. Similarly, in the many-species case the total density of
the majority species decays as $t^{-1}$ while the minorities densities
decay as $t^{-2}$.

We turn now to the determination of the cluster densities.
Introducing the generating functions,
$$
A(z,t)=\sum_{j=1}^\infty z^ja_j(t), \quad
B(z,t)=\sum_{j=1}^\infty z^jb_j(t),
\eqnoi
$$
transforms the governing equations into
$$
\dot A=A^2-2A(a+b), \quad
\dot B=B^2-2B(a+b).
\eqnoi
$$
These Bernoulli equations are straightforwardly solved to get
$$
A(z,t)={A_0(z)E(t)\over 1- A_0(z)\int_0^t dt'E(t')}, \quad
B(z,t)={B_0(z)E(t)\over 1- B_0(z)\int_0^t dt'E(t')},
\eqnoi
$$
with the shorthand notations $A_0(z)\equiv A(z,t=0), ~B_0(z)\equiv
B(z,t=0)$, and $E(t)\equiv\exp\left(-2\int_0^t
dt'[a(t')+b(t')]\right)$.

\last\ represents the general solution for arbitrary initial
conditions. Consider now the simplest but important case of
monodisperse, generally asymmetric, initial conditions
$$
a_k(0)=\delta_{k1}, \quad
b_k(0)=\l\delta_{k1}.
\eqnoi
$$
These initial conditions imply $A_0(z)=z, B_0(z)=\l z$.
Expansion of the resulting generating functions
$A(z,t)=zE(t)[1- z\int_0^t dt'E(t')]^{-1}$ and
$B(z,t)=\l zE(t)[1- z\l\int_0^t dt'E(t')]^{-1}$ yields
$$
a_k(t)=E(t)\left[\int_0^t dt'E(t')\right]^{k-1}, \quad
b_k(t)=\l^k E(t)\left[\int_0^t dt'E(t')\right]^{k-1}.
\eqnoi
$$

In the symmetric case $\l=1$ we easily compute
$E(t)=(1+3t)^{-4/3}$ and as a result
$$
a_k(t)=b_k(t)=(1+3t)^{-4/3}\left[1-(1+3t)^{-1/3}\right]^{k-1}.
\eqnoi
$$
Asymptotically, the cluster-mass distribution approaches the
scaling form
$$
a_k(t) \sim t^{-4/3}\exp(-x), \quad x\sim k t^{-1/3}.
\eqnoi
$$
The total mass densities, $m_a(t)=\sum_{k\geq 1}k a_k(t)$ and
$m_b(t)=\sum_{k\geq 1}k b_k(t)$, decrease with time,
$m_a=m_b\sim t^{-2/3}$. Note also that the quantity
$I\equiv m_a a^{-2/3}$  is conserved by the dynamics of
the aggregation-annihilation model subject to arbitrary {\it symmetric}
initial conditions. The conservation of $I$ is verified by a direct
computation which makes use of the evolution equations, $\dot a=-3a^2$
and $\dot m_a=-2a m_a$.  For pure aggregation processes,
the total mass is conserved. Thus the quantity  $I$ plays
the role of a ``mass'' in the aggregation-annihilation model.
Unfortunately, we could not understand the physical meaning
of the conservation law $I={\rm const}$.

In the asymmetric case we shall set $\l < 1$ thus forcing $B$-species
to be a minority. Making use of Eq.~(3) one can express
$E(t)=\exp\left(-2\int_0^t dt'[a(t')+b(t')]\right)$ as an explicit
function of $a$ and $b$, $E=(a(0)^{-1}-b(0)^{-1})/(a^{-1}-b^{-1})$.
In the long-time limit we use the asymptotic values,
$a\simeq t^{-1}$ and $b\simeq \l(1-\l)^{-3}t^{-2}$, to compute
$E(t)\simeq \l^{-1}(1-\l)b\simeq (1-\l)^{-2}t^{-2}$. Therefore,
$$
a_k(t) \simeq (1-\l)^{-2}t^{-2}\exp(-y), \quad
b_k(t) = \l^k a_k(t), \quad
y={k\over (1-\l)^2t}.
\eqnoi
$$
Thus, for arbitrary initial conditions the majority species
cluster-mass distribution can be written in the scaling form
$$
a_k(t) \sim t^{-w}\Phi[k/S(t)], \quad
S(t)\sim t^z,
\eqnoi
$$
where $S(t)$ is the characteristic mass. The scaling function is
exponential, $\Phi(x)=\exp(-x)$, both for symmetric and asymmetric
initial conditions.  In the symmetric case, the governing exponents
are $w=4/3$ and $z=1/3$.  The asymmetric case is equivalent to single
species aggregation, and the exponents are $w=2$ and $z=1$.  The
minority, on the other hand, does not scale according to the usual
definition although it can be expressed in the modified scaling form,
$b_k(t) \sim \l^k t^{-w}\Phi[k/S(t)]$. The modified scaling form also
indicates that two different mass scales are associated with the
minority species.  A growing scales $S(t)\sim t$, which is forced by
the majority species, and a time independent scale $S_{\l}=1/(1-\l)$
which dominates in the long-time limit.  The latter scale diverges,
$S_{\l}\cong 1/(1-\l)$, in the limit $\l\to 1$.  It is also
instructive to compute the total mass densities. By summing \back1\ we
find that as $t\to\infty$,
$$
m_a(t)\to (1-\l)^2, \quad
m_b(t)\to \l(1-\l)^{-4}t^{-2}.
\eqnoi
$$
Thus the final mass difference $\Delta m_{\infty} \equiv
m_a(\infty)-m_b(\infty)=m_a(\infty)$ may be expressed through the
initial mass difference $\Delta m_0 \equiv m_a(0)-m_b(0)=1-\l$ via a
surprisingly simple relation, $\Delta m_{\infty}=(\Delta m_0)^2$.  In
comparision, for the pure annihilation process, $A_i+B_j\to 0$, the
final mass density is significantly larger, $\Delta
m_{\infty}=\Delta m_0$.  More generally, the concentration difference,
$c_-=a-b$, is a conserved variable in the annihilation process, as
can be seen from the evolution equations $\dot a=\dot b=-2ab$. In the
aggregation-annihilation process a related ``hidden'' conservation law
exists. From the rate equations, one can verify that the quantity
$\tilde c_-= m_am_b c_-/ab$ is conserved. This unusual conservation
law is trivially satisfied in the case of equal initial densities,
where the aforementioned aggregation-like conservation law $m_a
a^{-2/3}={\rm const}$ holds.  We conclude that the
aggregation-annihilation process exhibits two nontrivial conservation
laws which are generalization to the usual conservation laws that
underly aggregation and annihilation separately.

We turn now to the general $n$-species model.  For asymmetric initial
conditions the majority species scales as in single-species
aggregation, while the minorities scale only in a modified sense.  For
the symmetric initial conditions, the cluster-mass distribution
approaches the scaling form of \back1\ with
$$
w_n={2n\over2n-1},\qquad{\rm and}\qquad z_n={1\over2n-1}.
\eqnoi
$$
Thus the exponents depend on the number of species $n$.  When $n=1$,
single-species aggregations is recovered $w_1=2$ and $z_1=1$, while
the limit $n\to\infty$, corresponds to single-species annihilation,
$w=1$ and $z=0$.  Denoting by $a^i$ ($m^i$) the concentration (mass)
of the $i^{\rm th}$ species, then the quantity, $M(a^i-a^j)/a^ia^j$ is
conserved for every $i\ne j$ independent of the initial conditions.
The factor $M$ is given by $M^{n-1}=\Pi_{i=1}^n a^i$.  Only $n-1$ of
these conservation laws are independent, similar to the $n-1$
independent conservation laws that unedrly the pure annihilation
process, $a^i-a^j={\rm const}$, for $i\ne j$.  For the case of
symmetric initial conditions, the aggregation induced conservation
laws read $m^i (a^i)^{-(2n-2)/(2n-1)}={\rm const}$ for $i=1
,\ldots,n$.

Generally, we expect that the exponents characterizing the scaling
behavior are universal, \ie, they depend only on important aspects of the
kinetics. The fact that the exponents do depend on the number of
species does not contradict universality.  However, the exponents in
the aggregation-annihilation system may depend also on the reaction
rates. To demonstrate that we consider the general case where the
aggregation rate and the annihilation rate are different.  We set the
aggregation rate to unity as previously and denote the annihilation
rate by $J$. In the $n$-species case with symmetric monodisperse
initial conditions we have
$$
\dot a_k=\sum_{i+j=k} a_ia_j-2a_k[1+(n-1)J]a, \quad
a_k(t=0)=\delta_{k1}.
\eqnoi
$$
By employing the above technique  we solve \last\
and find
$$
a_k(t)=(1+\nu t)^{-1-1/\nu}\left[1-(1+\nu t)^{-1/\nu}\right]^{k-1},
\eqnoi
$$
where the notation $\nu=1+2(n-1)J$ has been used. The
cluster mass distribution obeys the scaling form of Eq.~(18) with
the exponents
$$
w_n=2{1+(n-1)J\over 1+2(n-1)J}\qquad{\rm and}\qquad
z_n={1\over 1+2(n-1)J}.
\eqnoi
$$
Interestingly, both exponents dependent continuously on the (relative)
magnitude of the annihilation rate, $J$. This nonuniversality
contrasts the bulk of previously investigated aggregation models
\cite{ernst} (see, however, \cite{lr,paul1}). Varying the rate ratio
represents another way of interpolating between aggregation and
annihilation. Indeed, independent of $n$, when $J=0$, single-species
aggregation is recovered, while the limit $J\to\infty$ corresponds to
single-species annihilation.

We now consider aggregation-annihilation in the diffusion-controlled
limit in low dimensions. We shall study the particle coalescence model
(PCM) in which clusters occupy single lattice sites, hop to
nearest neighbor cites with a rate independent of the mass, and
aggregate or annihilate instantaneously whenever they meet. The PCM
allows one to focus on the kinetic aspects of the process and is well
suited for numerical implementation. Additionally, in the mean-field
approximation the PCM is governed by the same \back{18} which has been
examined previously.

Let us ignore the mass and the identity of the clusters and denote an
arbitrary cluster by $C$.  Then the reduced reaction process can be
described by the reaction scheme $C+C\to C$ and $C+C\to 0$ for
aggregation and annihilation, respectively. For both processes,
the total density $c=\sum_{k\geq 1} c_k$ is inversely proportional
to $N(t)$, the number of {\it distinct} sites visited by a single
random walk in $d$ dimensions \cite{zel}.
$N(t)$ is well known in probability theory \cite{feller}
and thus the density $c, ~c\sim N^{-1}$, behaves as
$$
c(t)\sim\cases{ t^{-1/2}&$d=1$;\cr
                t^{-1}\log(t) &$d=2$;\cr
                t^{-1}&$d>2$.\cr}
\eqnoi
$$
In other words, the time variable $t$ is replaced with a modified
time variable $N(t)$.  We further assume that the rate equation theory
describes the process in low dimensions, with the time variable
$N(t)$.  Although this approach is a heuristic one, it provides a
good approximation for the subcritical behavior.  For the symmetric
initial conditions, the characteristic mass for the case $n=2$ is given by
$$
S(t)\sim\cases{ t^{1/6}&$d=1$;\cr
                \left(t/\log(t)\right)^{1/3} &$d=2$;\cr
                t^{1/3}&$d>2$.\cr}
\eqnoi
$$
Using the asymptotic forms of the concentration and the typical mass,
a scaling form similar to the one of Eq.~(18) can be written.
The same analysis can be repeated for the general multi-species case,
and we quote the scaling exponents in one dimension only,
$$
w_n={n\over2n-1},\qquad{\rm and}\qquad z_n={1\over2(2n-1)}.
\eqnoi
$$
In the extreme cases of $n=1$ and $n\to\infty$, these exponents agree
with the solutions to single-species aggregation and single-species
annihilation, respectively \cite{spoug}.
In the case of asymmetric
initial conditions, the majority species concentration decays
according to single-species aggregation, or equivalently, according
to Eq.~(23).  The minority species, on the other hand, decays much
faster. Using the above heuristic argument, we find that in the
long-time limit, the minority-species concentration decays as
$t^{-1}$ in 1D, and as $t^{-2}\log^2(t)$ in 2D.

It is interesting to compare the above theoretical predictions with
numerical simulations. We have performed simulations for $d=1, 2$, and 3.
The numerical implementation of the process is simple. Initially all
$L^d$ sites on the cubic lattice are occupied with monomers. An
elemental simulation step consists of picking a cluster at random and
moving it to a randomly chosen neighboring site. If the site is
occupied, an aggregation or an annihilation event takes place,
depending on the identity of the two clusters. Time is updated by the
inverse of the total number of particles in the system after each
step. The linear dimension of the lattice used in the simulation was
$L=3\times10^7$, $4\times10^3$, $2\times 10^2$ in 1D, 2D, and 3D
respectively.  The corresponding number of particles was thus
$30\times10^6$, $16\times10^6$, and $8\times 10^6$ for 1D, 2D, and 3D.

We discuss first results for the symmetric initial conditions
involving two species. To verify that the process belongs to the PCM
universality class we measured the density of particles as a function
of time. Indeed, \back2\ appears to hold asymptotically (see Figure
1).  We also measured the average cluster size $\langle
k(t)\rangle=\sum k c_k/\sum c_k$. We expect that the average cluster
size is proportional to the characteristic size $S(t)$,
asymptotically.  For $d=3$, the mean-field prediction, $\langle
k(t)\rangle \sim t^{1/3}$ is verified (see Figure 2). Note that the
simulation results are reliable up to time $\propto L^2$ due to finite
size effects. At the critical dimension, $d_c=2$, the average mass
grows slightly slower, consistent with the logarithmic correction of
\back1. In 1D, the growth exponent $z$ as determined by a least square
fit is $z=0.190\pm 0.002$. However, the corresponding value from the
approximate theory is lower $z=1/6$. The simulation was carried over a
relatively large temporal range, suggesting that the growth exponent
is indeed different than suggested by the heuristic theory.  A similar
trend is observed when more than two species are involved. We find
that the numerically determined exponents $z_3\cong0.12$ and
$z_4\cong0.09$ are slightly higher than their theoretical counterparts
$z_3=1/10$ and $z_4=1/14$. We have made some additional consistency
checks. For example, we verified that the cluster mass distribution
follows the scaling form of \back8.  Above the critical dimension the
scaling function is indeed a simple exponential, in agreement with the
rate equation predictions.

In summary, we introduced an annihilation-aggregation model and
presented the solution to the governing rate equations. The mass
distribution follows a general scaling form with nontrivial growth
exponents, and the process is characterized by unusual conservation
laws.  The kinetic behavior depends both on the number of species and
on the specific rates of the annihilation and the aggregation
processes. Numerical simulations confirm the rate equation predictions
above the critical dimension. Below the critical dimension, a
heuristic argument provides a good estimate for the mass distribution.
Nontrivial exponents describe the process in one-dimension and it
would be interesting to study this system rigorously. Furthermore, the
underlying conservation laws in low dimensions are especially
intriguing.


\bigskip\centerline{\bf Acknowledgments}\medskip

E.B. was supported in part by the MRSEC
Program of the National Science Foundation under Award Number
DMR-9400379 and by NSF under Award Number 92-08527.
P.L.K. was supported by ARO grant \#DAAH04-93-G-0021
and NSF grant \#DMR-9219845.

\bigskip\centerline{\bf Figure Captions}\medskip

\figi\ The total density versus time in 1D, 2D and 3D.  Shown are the
simulation data (bullets) and the theoretical prediction of
Eq.~(24) (solid line).

\figi\ The average cluster size $\langle k(t)\rangle$ versus $t$ in
1D, 2D, and 3D. Shown are simulation data (bullets) and lines of
slope $0.19$ (solid line) and $1/3$ (dashed line) for reference.


\begin{references}

\bibitem{smoke} S.~K.~Frielander, {\sl Smoke, Dust and Haze:
                Fundamentals of Aerosol Behavior} (Wiley, New York, 1977).

\bibitem{meak}  P.~Meakin, \rpp 55 157 1992 .

\bibitem{smol}  M.~V.~Smoluchowski, \zpc 92 215 1917 .

\bibitem{ernst} M.~H.~Ernst,
                in {\sl Fundamental Problems in Statistical Physics VI},
                ed. E.~G.~D.~Cohen (Elsevier, New York, 1985).

\bibitem{zel}   Ya.~B.~Zel'dovich and A.~S.~Mikhailov, \usp 30 23 1988 .

\bibitem{bAR}   D.~ben-Avraham and S.~Redner, \pra 34 501 1986 .

\bibitem{leb}   M. Bramson and J. L. Lebowitz, \prl 61 2397 1988 .

\bibitem{vD}    P.~G.~J.~van Dongen, \prl 63 1281 1989 .

\bibitem{kang}  K.~Kang and S.~Redner, \prl 52 955 1984 .

\bibitem{paul1} P.~L.~Krapivsky, \pA 198 135 1993 .

\bibitem{paul2} P.~L.~Krapivsky, \pA 198 150 1993 ;
                \pA 198 157 1993 .

\bibitem{blum}  I.~M.~Sokolov and A.~Blumen, \pre 50 2335 1994 .

\bibitem{lr}    F.~Leyvraz and S.~Redner, \pra 36 4033 1987 .

\bibitem{feller} W.~Feller, {\it An Introduction to Probability
        Theory, Vols.~1 and 2} (Wiley, New York, 1971).

\bibitem{spoug} J.~L.~Spouge, \prl 60 871 1988 ;
 A.~A.~Lushnikov, \jetp 64 811 1986 ;
Z.~Cheng, S.~Redner, and F.~Leyvraz, \prl 62 2321 1989 ;
D.~ben-Avraham, M.~A.~Burschka, and C.~R.~Doering, \jsp
60 695 1990 .






\end{references}
\end{document}